\documentclass[aps,preprintnumbers,nofootinbibt,referee, superscriptaddress, keywords]{revtex4}
\usepackage{graphicx}
\usepackage{amsmath}
\usepackage{bm}
\usepackage{color}
\usepackage{mathptmx}

\def\be{\begin{equation}}
\def\ee{\end{equation}}
\def\bea{\begin{eqnarray}}
\def\eea{\end{eqnarray}}

\begin{document}

\title{Series solution of the Susceptible-Infected-Recovered (SIR) epidemic
model with vital dynamics via the Adomian and Laplace-Adomian Decomposition Methods}
\author{Tiberiu Harko}
\email{tiberiu.harko@aira.astro.ro}
\affiliation{Astronomical Observatory, 19 Ciresilor Street, 400487 Cluj-Napoca, Romania,}
\affiliation{Faculty of Physics, Babes-Bolyai University, 1 Kogalniceanu Street, 400084
Cluj-Napoca, Romania}
\affiliation{School of Physics, Sun Yat-Sen University, Xingang Road, 510275 Guangzhou,
People's Republic of China}
\author{Man Kwong Mak}
\affiliation{Departamento de Fisica, Facultad de Ciencias Naturales, Universidad de
Atacama, Copayapu 485, Copiapo, Chile}
\email{mankwongmak@gmail.com}

\begin{abstract}
The Susceptible-Infected-Recovered (SIR) epidemic model as well as its generalizations are extensively used
for the study of the spread of infectious diseases, and for the understanding of the dynamical evolution of epidemics. From SIR type models only the model without vital dynamics has an exact analytic solution, which can be obtained in an exact parametric form. The SIR model with vital dynamics, the simplest extension of the basic SIR model, does not admit a closed form representation of the solution. However, in order to perform the comparison with the epidemiological data  accurate representations of the time evolution of the SIR  model with vital dynamics would
be very useful. In the present paper, we obtain first the basic evolution equation of the SIR model with vital dynamics, which is given by a strongly nonlinear second order differential equation. Then we obtain a series representation of the solution of the model, by using the Adomian and Laplace-Adomian Decomposition Methods to solve the dynamical evolution equation of the model. The solutions are
expressed in the form of infinite series. The series representations of the
time evolution of the SIR model with vital dynamics are compared with the exact numerical
solutions of the model, and we find that, at least for a specific range of parameters, there is a good agreement between the
Adomian and Laplace-Adomian semianalytical solutions, containing only a small number of terms, and
the numerical results.

\end{abstract}

\maketitle
\tableofcontents

%\pacs{67.85.Jk, 04.40.Dg, 95.30.Cq, 95.30.Sf}

\section{Introduction}

The study of the epidemic mathematical models in different formal formulation has been proved to be of crucial interest in the understanding of the dynamics, spread and control of epidemic diseases \cite{1a,Mur,2a,3a,4a,5a}. The simplest of the epidemic models are the so-called deterministic
compartmental models, consisting of at least three compartments: given by: the number of susceptible individuals (S),  the number of infectious individuals (I), and the number of removed (and immune), or deceased individuals (R), respectively. The first compartmental epidemiological models have been proposed and investigated in \cite{2}. Despite their apparent phenomenological simplicity, the mathematical equations describing the SIR type models are essentially nonlinear, with this nonlinearity raising a number of important and interesting mathematical problems in the study of even the simplest models of epidemics. On the other hand the SIR type models still have a powerful predictive and investigating power,
and many of them have been used to investigate the recent COVID-19
pandemic \cite{China}, like, for example, in \cite{r1,r2,r3,r3a,r4,r5,r6,r7,r8,r9,r10,r11,r12,r13,r14, r15}.

The basic equations of the SIR model are given by the following nonlinear system of ordinary differential equations \cite{2,Harko1}
\begin{equation}\label{A3}
\frac{dx}{dt}=-\beta x\left( t\right) y\left( t\right) ,
\frac{dy}{dt}=\beta x\left( t\right) y\left( t\right) -\gamma y\left(
t\right) ,
\frac{dz}{dt}=\gamma y\left( t\right) ,
\end{equation}%
respectively, where $x(t)>0$, $y(t)>0$ and $z(t)>0$, $\forall t\geq 0$. The system of equations (\ref{A3}) must be integrated with the initial conditions $x\left( 0\right) =N_{1}\geq 0$, $%
y\left( 0\right) =N_{2}\geq 0$ and $z\left( 0\right) =N_{3}\geq 0$, respectively, where $%
N_{i}\in \Re $, $i=1,2,3$. The time evolution of the SIR epidemic model, as well as its intrinsic dynamics is determined by two epidemiological parameters only, the infection rate $\beta $, and the mean recovery rate $\gamma $, respectively, which are assumed to be positive constants. The exact solution of the SIR model (\ref{A3}) can be obtained in an exact parametric form, and it is given by \cite{Harko1}
\be
x=x_0u, y=N+\frac{\gamma}{\beta} \ln u -x_0u, z=-\frac{\gamma}{\beta}\ln u, t-t_0=\int_{u_0}^u{\frac{d\xi }{\xi\left(x_0\beta \ln \xi-\gamma \xi-N\right)}},
\ee
where $N=\sum _{i=1}^3{N_i}$, $x_0=N_1e^{(\beta /\gamma)}N_3$, and $u$ is a parameter.  In these equations $x (t)$ gives the number of individuals not yet infected with the epidemics at time $t$, or those susceptible to the infection, $y (t)$ represents the number of individuals who have already been infected with the disease, and hence are able to spread the disease to the persons in the susceptible category, while
$z (t)$ denotes the individuals who have been infected, but did recover from the disease.  For a pedagogical discussion of the SIR model and of its exact solution see \cite{Ped}. However, due to the integral representation of the time variable, some alternative representations and methods have been used for the study of the solution of the SIR model \cite{6,7,8}, including the variational iteration method, and the homotopy perturbation method.

A powerful method for obtaining semianalytical solutions of strongly nonlinear differential equations is represented by the Adomian Decomposition Method \cite{Ad1,Ad2,Ad3,Ad4}, whose basic idea is to decompose the nonlinear terms appearing in a differential equation in terms of the Adomian polynomials, which are constructed recursively.   The Adomian Decomposition Method, as well as its very effective version, the Laplace-Adomian Decomposition Method, has been extensively used to study the approximate semianalytical solutions of different types of differential equations, and physical models \cite{Ad5,Ad6,Ad7,Ad8,Ad9,Ad10}. The Adomian Decomposition Method was used for the study of the SIR model in \cite{5}, while a series solution of the SIR model by using the Laplace-Adomian Decomposition Method was obtained in \cite{LapH}, where the solutions have been expressed in the form of infinite series. The series representations of the time evolution of the SIR compartments obtained in \cite{LapH} have been compared with the exact numerical solutions of the model, and, for a specific range of the model parameters, a good agreement between the Laplace-Adomian semianalytical solutions, containing only three terms, and the numerical results, was obtained.

One of the simplest generalizations of the SIR model is given by the SIR model with vital dynamics \cite{1a,Mur,3a,5a}, in which the number of deaths and births is included in the model via a new constant $\mu$. Usually it is assumed that the death rate is equal to the birth rate. Even if the mathematical modifications of the initial SIR model with $\mu =0$ look, at first sight, minimal, the SIR model with vital dynamics has very important differences as compared to the SIR model (\ref{A3}). The SIR model without vital dynamics with $\mu =0$, has a first integral $N = S + I + R$, which
is the total number of individuals in the given population. Moreover, it also has a second first integral $G(S,I,R)=\beta R+\gamma \ln S $ \cite{Chen}. Hence the SIR system with $\mu = 0$ is a completely integrable system with two functionally independent first integrals. On the other hand in the case of $\mu \neq 0$, in \cite{Chen} it was
proved that the SIR model has no polynomial or proper rational first integrals. This result is obtained by studying the invariant algebraic surfaces.
Moreover, although the SIR model with $\mu \neq  0$ is not integrable, and hence it does not have an exact solution, the global dynamics of the SIR model with vital dynamics  can be studied based on the existence of an invariant algebraic surface.

In \cite{Harko1} it was shown that the generalization of the SIR model with $\mu \neq 0$, including births and deaths, and described by a nonlinear system of differential equations, can be reduced to an Abel type equation. The reduction of the SIR model with vital dynamics to an Abel type first order differential equation  greatly simplifies the analysis of its properties. An approximate solution of the Abel equation was obtained by using a perturbative approach, in a power series form. Moreover, the general solution of the SIR model with vital dynamics can be represented in an exact parametric form.

In the present work we consider the possibility of obtaining some accurate semianalytical solutions of the equations of the SIR model with vital dynamics by using the Adomian and the Laplace-Adomian Decomposition Methods, respectively. As a first step in our analysis we reduce the SIR model with $\mu \neq 0$ to a basic second order differential equation, describing the evolution of the individuals infected with the disease. Then, once the solution of the basic equation is known, the general solution of the SIR model with vital dynamics can be obtained in terms of the variable $u$ related to the number $y$ of the infected individuals.

In order to obtain some approximate solutions of the basic evolution equation we will apply to it both the Adomian and the Laplace-Adomian Decomposition Methods, We obtain in each case the recurrence relations giving the successive terms in the Adomian series representation as a function of the Adomian polynomials. In the case of the Adomian Decomposition Method the iterative solution can be evaluated exactly only for the first two terms of the series expansion, while for the Laplace-Adomian Decomposition Method all the terms in the series expansion can be obtained exactly. We also perform a careful comparison of the semianalytical results with the exact numerical solutions, and it turns out that for certain ranges of the model parameters $\beta, \gamma, \mu)$ both the Adomian and the Laplace-Adomian Decomposition Methods can give a good description of the numerical results.

The present paper is organized as follows. The basic equation describing the dynamical evolution of the SIR model with vital dynamics is obtained in Section~\ref{II}. The semianalytical solutions of the SIR system for $\mu \neq 0$ are obtained, by using the Adomian and the Laplace-Adomian Decomposition Methods in Section~\ref{III}. The comparison with the exact numerical solutions is performed in Section~\ref{IV}. We discuss and conclude our results in Section~\ref{V}.

\section{The basic evolution equations of the SIR model with vital
dynamics}\label{II}

In the present Section we will obtain the basic equation describing the dynamics  of the SIR model with vital dynamics. Then, by using this equation,  we will obtain semianalytical, but still having a high numerical precision, solutions of the SIR model with vital dynamics, represented in the forms of  Adomian type series, containing exponential terms.

The strongly nonlinear system of equations describing the SIR model with vital dynamics is given by \cite{1a,Mur,3a,5a},
\begin{equation}
\frac{dx}{dt}=-\frac{\beta}{N} xy+\mu \left( N-x\right) ,  \label{K1}
\end{equation}%
\begin{equation}
\frac{dy}{dt}=\frac{\beta}{N} xy-\left(\gamma +\mu \right)y,  \label{K2}
\end{equation}%
\begin{equation}
\frac{dz}{dt}=\gamma y-\mu z,  \label{K3}
\end{equation}%
where $x(t)>0$, $y(t)>0$ and $z(t)>0$, $\forall t\geq 0$. The system of
strongly nonlinear differential equations (\ref{K1})-(\ref{K3}) must be
integrated with the initial conditions $x(0)=N_1\geq 0$, $y(0)=N_2\geq 0$
and $z(0)=N_3\geq 0$, with the constants $N_i$, $i=1,2,3$ satisfying the
condition $\sum _{i=1}^3{N_i}=N$. The time evolution of the model is
determined by three epidemiological parameters, the infection rate $\beta $,
the mean recovery rate $\gamma $, and $\mu$, representing the natural death
rate, which in the present investigation is assumed to be equal to the birth
rate. In the following $\beta $, $\gamma $, and $\mu$ are assumed to be
positive constants.

By adding Eqs.~(\ref{K1})--(\ref{K3}), and integrating the resulting
equation we immediately obtain
\begin{equation}
x(t)+y(t)+z(t)=N+N_{0}e^{-\mu t},
\end{equation}
where $N_{0}$ is an arbitrary integration constant. In order to assure that
the total number of individuals in the group is a constant,
\begin{equation}
x(t)+y(t)+z(t)=N,\forall t\geq 0,
\end{equation}
we must fix the integration constant $N_0$ as zero, $N_{0}=0$. In the
following, in order to significantly simplify the mathematical formalism we
introduce first a new function $u(t)$, related to $y(t)$ by
\begin{equation}  \label{6}
y(t)=N_2e^{u(t)},
\end{equation}
and which satisfies the initial condition
\begin{equation}
u(0)=0,
\end{equation}
and thus obviously giving $y(0)=N_2$.

Then from Eq.~(\ref{K2}) we obtain for $x(t)$ the simple expression
\begin{equation}  \label{7}
x(t)=\frac{1}{\left(\beta/N\right)}\frac{du(t)}{dt}+\frac{\gamma +\mu}{%
\left(\beta /N\right)}.
\end{equation}

By substituting Eqs.~(\ref{6}) and (\ref{7}) into Eq.~(\ref{K1}), the latter
becomes
\begin{equation}  \label{8}
\frac{d^2u(t)}{dt^2}+\mu \frac{du(t)}{dt}+\mu \left[\left(\gamma
+\mu\right)-\beta\right]=-\beta \frac{N_2}{N}\left[\frac{du(t)}{dt}+(\gamma
+\mu)\right]e^{u(t)}.
\end{equation}

Eq.~(\ref{8}) can be reformulated as
\begin{equation}  \label{9}
\frac{d^2u(t)}{dt^2}+\mu \frac{du(t)}{dt}+\mu \left[\left(\gamma
+\mu\right)-\beta \right]=-\beta \frac{N_2}{N}\frac{d}{dt}e^{u(t)}-\beta
\frac{N_2}{N}(\gamma +\mu)e^{u(t)}.
\end{equation}

Eq.~(\ref{9}) represents the basic dynamical evolution equation of the SIR
model with vital dynamics. It must be solved with the initial conditions
\begin{equation}
u(0)=0, \left.\frac{du(t)}{dt}\right|_{t=0}=\beta \frac{N_1}{N}-(\gamma
+\mu).
\end{equation}

By representing $z(t)$ as
\begin{equation}
z(t)=e^{-\mu t}w(t),
\end{equation}
Eq.~(\ref{K3}) becomes
\begin{equation}
\frac{dw(t)}{dt}=\gamma N_2e^{\mu t+u(t)},
\end{equation}
giving
\begin{equation}
w(t)=w(0)+\gamma N_2\int_0^t{\ e^{\mu t+u(t)}dt},
\end{equation}
and
\begin{equation}
z(t)= e^{-\mu t}\left[N_3+\gamma N_2\int_0^t{\ e^{\mu t+u(t)}dt}\right],
\end{equation}
respectively, where we have used the condition $w(0)=z(0)=N_3$.

\section{The Adomian and the Laplace - Adomian Decomposition Methods}\label{III}

The general solution of Eq.~(\ref{9}) cannot be obtained in a closed (exact)
form. In the present Section we will consider the applications of the
Adomian Decomposition Method for the study of the SIR model with vital
dynamics. In order to obtain a semianalytical solution of the model we will
consider first the solution of the basic evolution equation (\ref{9})by
obtaining the Adomian type recursive relations for solving Eq.~(\ref{9}) in
both the standard Adomian as well as in the Laplace-Adomian Decomposition
methods.

\subsection{The Adomian Decomposition Method}

We will begin our investigations of Eq.~(\ref{9}) by applying first the
standard Adomian Decomposition Method. We integrate Eq. (\ref{9}) between 0
and $t$. With the use of the initial conditions we obtain
\begin{equation}
\frac{du(t)}{dt}-\left[ \beta \frac{N_{1}}{N}-(\gamma +\mu )\right] +\mu
u(t)+\mu \left[ \left( \gamma +\mu \right) - \beta\right] t=-\beta \frac{%
N_{2}}{N}e^{u(t)}+\beta \frac{N_{2}}{N}-\beta \frac{N_{2}}{N}(\gamma +\mu
)\int_{0}^{t}e^{u(t)}dt.  \label{10a}
\end{equation}

Again we integrate Eq. (\ref{10a}) between 0 and $t$ to find
\begin{equation}
u(t)=\left[ \beta \frac{\left( N_{1}+N_{2}\right)}{N} -(\gamma +\mu )\right]
t-\frac{1}{2}\mu \left[ \left( \gamma +\mu \right) -\beta \right] t^{2}-\mu
\int_{0}^{t}u(t)dt-\beta \frac{N_{2}}{N}\int_{0}^{t}e^{u(t)}dt-\beta \frac{%
N_{2}}{N}(\gamma +\mu )\int_{0}^{t}\int_{0}^{w}e^{u(s)}dsdw.  \label{11b}
\end{equation}

Now we will apply the Adomian Decomposition method to Eq. (\ref{11b}). For
this we assume
\begin{equation}
u(t)=\sum_{n=0}^{\infty }u_{n}(t),  \label{3.5}
\end{equation}%
and
\begin{equation}
e^{u(t)}=\sum_{n=0}^{\infty }A_{n},  \label{3.6}
\end{equation}
where $A_{n}(t)$ are the Adomian polynomials, defined for an arbitrary
function $f(u(t))$ according to the general formula \cite{Ad3}
\begin{equation}  \label{adom}
A_{n}=\left. \frac{1}{n!}\frac{d^{n}}{d\lambda ^{n}}f\left(
\sum_{i=0}^{\infty }{\lambda ^{i}u_{i}}\right) \right\vert _{\lambda =0}.
\end{equation}

Substituting Eqs. (\ref{3.5}) and (\ref{3.6}) into Eq. (\ref{11b}) we obtain
\begin{eqnarray}
\sum_{n=0}^{\infty }u_{n}(t)&=&\left[ \beta \frac{\left( N_{1}+N_{2}\right)}{%
N} -(\gamma +\mu )\right] t-\frac{1}{2}\mu \left[ \left( \gamma +\mu \right)
-\beta \right] t^{2}-\mu \sum_{n=0}^{\infty }\int _0^t{u_{n}(t)dt}-\beta
\frac{N_{2}}{N}\sum_{n=0}^{\infty }\int_{0}^{t}{A_{n}(t)dt}-  \notag \\
&&\beta \frac{N_{2}}{N}(\gamma +\mu )\sum_{n=0}^{\infty
}\int_{0}^{t}\int_{0}^{w}{A_{n}\left( s\right) dsdw}.  \label{11c}
\end{eqnarray}

Explicitly, we can write the above equation as
\begin{eqnarray}
u_{0}(t)+u_{1}(t)+u_{2}(t)+... &=&\left[ \beta \frac{\left(
N_{1}+N_{2}\right)}{N} -(\gamma +\mu )\right] t-\frac{1}{2}\mu \left[ \left(
\gamma +\mu \right) -\beta \right] t^{2}-  \notag \\
&&\mu \left[ \int_{0}^{t}u_{0}(t)dt+\int_{0}^{t}u_{1}(t)dt+%
\int_{0}^{t}u_{2}(t)dt+...\right] - \nonumber\\
&&\beta \frac{N_{2}}{N}\left[ \int_{0}^{t}A_{0}(t)dt+\int_{0}^{t}A_{1}(t)dt+%
\int_{0}^{t}A_{2}(t)dt+...\right] -  \notag \\
&&\beta \frac{N_{2}}{N}(\gamma +\mu )\left[ \int_{0}^{t}\int_{0}^{w}A_{0}%
\left( s\right) dsdw+\int_{0}^{t}\int_{0}^{w}A_{1}\left( s\right)
dsdw+\int_{0}^{t}\int_{0}^{w}A_{2}\left( s\right) dsdw+...\right] .
\end{eqnarray}

Now we take
\begin{equation}
u_{0}(t)=\left[ \beta \frac{\left( N_{1}+N_{2}\right)}{N} -(\gamma +\mu )%
\right] t-\frac{1}{2}\mu \left[ \left( \gamma +\mu \right) -\beta \right]
t^{2},
\end{equation}
\begin{equation}
u_{1}(t)=-\mu \int_{0}^{t}u_{0}(t)dt-\beta \frac{N_{2}}{N}%
\int_{0}^{t}A_{0}(t)dt-\beta \frac{N_{2}}{N}(\gamma +\mu
)\int_{0}^{t}\int_{0}^{w}A_{0}\left( s\right) dsdw,
\end{equation}

\begin{equation}
u_{2}(t)=-\mu \int_{0}^{t}u_{1}(t)dt-\beta \frac{N_{2}}{N}%
\int_{0}^{t}A_{1}(t)dt-\beta \frac{N_{2}}{N}(\gamma +\mu
)\int_{0}^{t}\int_{0}^{w}A_{1}\left( s\right) dsdw,
\end{equation}

\begin{equation}
...
\end{equation}

Therefore we have obtained the iteration
\begin{equation}  \label{13a}
u_{0}(t)=\left[ \beta \frac{\left( N_{1}+N_{2}\right)}{N} -(\gamma +\mu )%
\right] t-\frac{1}{2}\mu \left[ \left( \gamma +\mu \right) -\beta \right]
t^{2},
\end{equation}
\begin{equation}
u_{n}(t)=-\mu \int_{0}^{t}u_{n-1}(t)dt-\beta \frac{N_{2}}{N}%
\int_{0}^{t}A_{n-1}(w)dw-\beta \frac{N_{2}}{N}(\gamma +\mu
)\int_{0}^{t}\int_{0}^{w}A_{n-1}\left( s\right) dsdw,n\in \left[
1,2,3,...,\infty \right) .  \label{13b}
\end{equation}

Now we can use the Cauchy formula for repeated integration, which gives
\begin{equation}
\int_{0}^{t}\int_{0}^{w}A_{n-1}\left( s\right) dsdw=\int_{0}^{t}\left(
t-w\right) A_{n-1}\left( w\right) dw=t\int_{0}^{t}A_{n-1}\left( w\right)
dw-\int_{0}^{t}wA_{n-1}\left( w\right) dw.  \label{14a}
\end{equation}

Thus Eq. (\ref{13b}) becomes
\begin{equation}
u_{n}(t)=-\mu \int_{0}^{t}u_{n-1}(t)dt-\beta \frac{N_{2}}{N}\left[ 1+(\gamma
+\mu )t\right] \int_{0}^{t}A_{n-1}(w)dw+\beta \frac{N_{2}}{N}(\gamma +\mu
)\int_{0}^{t}wA_{n-1}\left( w\right) dw,n\in \left[ 1,2,3,...,\infty \right)
.  \label{15a}
\end{equation}

In the following we denote
\begin{equation}
a=(\gamma +\mu )-\beta \frac{\left( N_{1}+N_{2}\right) }{N},b=\frac{1}{2}\mu %
\left[ \left( \gamma +\mu \right) -\beta \right] .
\end{equation}

Hence
\begin{equation}
u_{0}(t)=-at-bt^{2}.
\end{equation}

The first Adomian polynomial is given by
\begin{equation}
A_{0}(t)=e^{u_{0}(t)}=e^{-at-bt^{2}}.
\end{equation}

For $u_{1}(t)$ we obtain
\bea
u_{1}(t)&=&\frac{3\sqrt{\pi }\beta N_{2}e^{\frac{a^{2}}{4b}}}{12b^{3/2}N}%
\Bigg\{ \left[ \text{erf}\left( \frac{a}{2\sqrt{b}}\right) -\text{erf}\left(
\frac{a+2bt}{2\sqrt{b}}\right) \right] \left[ a(\gamma +\mu )+2b(t(\gamma
+\mu )+1)\right] +\nonumber\\
&&2\sqrt{b}\left[ 3\beta N_{2}(\gamma +\mu )\left(
1-e^{-(a+bt)t}\right) +b\mu N(3a+2bt)t^{2}\right] \Bigg\} ,
\eea
where $\text{erf}(z)=\left(2/\sqrt{\pi}\right)\int_0^z{e^{-t^2}dt}$ is the error function. The next term $u_2$ cannot be obtained in an exact form, and hence we will not present it.
%\bea
%u_2(t)=-\frac{48 \beta  N_2 \left((t (\gamma +\mu )+1) \int_0^t \frac{e^{-w (a+b w)} \left(3
%   \sqrt{\pi } \beta  N_2 e^{\frac{a^2}{4 b}} \left(\text{erf}\left(\frac{a}{2
%   \sqrt{b}}\right)-\text{erf}\left(\frac{a+2 b w}{2 \sqrt{b}}\right)\right) (a (\gamma +\mu
%   )+2 b (w (\gamma +\mu )+1))+2 \sqrt{b} \left(3 \beta  N_2 (\gamma +\mu )
%   \left(1-e^{-w (a+b w)}\right)+b \mu  N w^2 (3 a+2 b w)\right)\right)}{12 b^{3/2}
%   N} \, dw-(\gamma +\mu ) \int_0^t \frac{w e^{-w (a+b w)} \left(3 \sqrt{\pi } \beta
%    N_2 e^{\frac{a^2}{4 b}} \left(\text{erf}\left(\frac{a}{2
%   \sqrt{b}}\right)-\text{erf}\left(\frac{a+2 b w}{2 \sqrt{b}}\right)\right) (a (\gamma +\mu
%   )+2 b (w (\gamma +\mu )+1))+2 \sqrt{b} \left(3 \beta  N_2 (\gamma +\mu )
%   \left(1-e^{-w (a+b w)}\right)+b \mu  N w^2 (3 a+2 b w)\right)\right)}{12 b^{3/2}
%   N} \, dw\right)+\frac{3 \sqrt{\pi } \beta  \mu  N_2 e^{\frac{a^2}{4 b}}
%   \left(\text{erf}\left(\frac{a}{2 \sqrt{b}}\right)-\text{erf}\left(\frac{a+2 b t}{2
%   \sqrt{b}}\right)\right) \left(a^2 (\gamma +\mu )+4 a b (t (\gamma +\mu )+1)+2 b (2 b t (t
%   (\gamma +\mu )+2)+\gamma +\mu )\right)}{b^{5/2}}+\frac{2 \mu  \left(4 a b^2 \mu
%   N t^3-3 \beta  N_2 e^{-t (a+b t)} (a (\gamma +\mu )+2 b (t (\gamma +\mu
%   )+2))+3 a \beta  N_2 (\gamma +\mu )+2 b^3 \mu  N t^4+12 b \beta  N_2
%   (t (\gamma +\mu )+1)\right)}{b^2}}{48 N}
%\eea

\subsection{Series solution of the basic equation of the SIR model with
vital dynamics via the Laplace-Adomian method}

In the following we will look for a series solution of Eq.~(\ref{9}), by
using the Laplace-Adomian Decomposition Method.

In the Laplace-Adomian Decomposition Method we first apply the Laplace
transformation operator $\mathcal{L}_x$, defined as $\mathcal{L}_x[f(x)](s)
= \int_0^{\infty}{f(x)e^{-sx}dx}$, to Eq.~(\ref{9}), thus obtaining
\begin{equation}
\mathcal{L}_t\left[\frac{d^2u(t)}{dt^2}\right](s) +\mu \mathcal{L}_t\left[%
\frac{du(t)}{dt}\right](s)+\mathcal{L}_t\left[\mu \left(\gamma
+\mu\right)-\beta \mu \right]=-\beta \frac{N_2}{N} \mathcal{L}_t\left[\frac{d%
}{dt}e^{u(t)}\right](s)-\beta \left(\gamma +\mu\right) \frac{N_2}{N}\mathcal{%
L}_t\left[e^{u(t)}\right](s).
\end{equation}

By using the properties of the Laplace transform, $\mathcal{L}_t \left[%
d^2f(t)/dt^2\right](s)=s^2\mathcal{L}_t\left[f(t)\right](s)-sf(0)-f^{\prime
}(0)$ and $\mathcal{L}_t \left[df(t)/dt\right](s)=s\mathcal{L}_t\left[f(t)%
\right](s)-f(0)$, we find
\begin{eqnarray}
&&s^2\mathcal{L}_t\left[u(t)\right](s)+(\gamma +\mu)-\beta \frac{N_1}{N}+\mu
s \mathcal{L}_t\left[u(t)\right](s)+\frac{\mu \left[(\gamma +\mu)-\beta %
\right] }{s}=  \notag \\
&&-\beta \frac{N_2}{N}s \mathcal{L}_t \left[e^{u(t)}\right](s)+\beta \frac{%
N_2}{N}-\beta \left(\gamma +\mu\right)\frac{N_2}{N}\mathcal{L}_t \left[%
e^{u(t)}\right](s),
\end{eqnarray}
giving for the Laplace transform of $u(t)$ the equation
\begin{equation}  \label{6a}
\mathcal{L}_t[u(t)](s)=\frac{\beta \left(N_1+N_2\right)/N-(\gamma +\mu)}{%
s(s+\mu)}-\frac{\mu \left[(\gamma +\mu)-\beta \right] }{s^2(s+\mu)}-\frac{%
\left(\beta/N\right)\left[s+(\gamma +\mu)\right]N_2}{s(s+\mu)}\mathcal{L}_t%
\left[e^{u(t)}\right](s).
\end{equation}
We assume now that the function $u(t)$ can be represented in the form of an
infinite series,
\begin{equation}  \label{7a}
u(t)=\sum_{n=0}^{\infty }u_{n}(t),
\end{equation}%
where all the terms $v_{n}(t)$, $n=0,1,2,...$ can be computed recursively.
As for the nonlinear operator $f(u(t))=e^{u(t)}$, we assume that it can be
decomposed as
\begin{equation}  \label{8a}
f(u(t))=e^{u(t)}=\sum_{n=0}^{\infty }A_{n}(t),
\end{equation}%
where $A_{n}(t)$ are the Adomian polynomials, defined according to Eq.~(\ref%
{adom}).

The first five Adomian polynomials can be obtained in the following form,
\begin{equation}
A_{0}=f\left( u_0\right) ,  \label{Ad0}
\end{equation}%
\begin{equation}
A_{1}=u_{1}f^{\prime }\left( u_0\right) ,  \label{Ad1}
\end{equation}%
\begin{equation}
A_{2}=u_{2}f^{\prime }\left( u_0\right) +\frac{1}{2}u_{1}^{2}f^{\prime
\prime }\left( u_0\right) ,  \label{Ad2}
\end{equation}%
\begin{equation}
A_{3}=u_{3}f^{\prime }\left( u_0\right) +u_{1}u_{2}f^{\prime \prime }\left(
u_0\right) +\frac{1}{6}u_{1}^{3}f^{\prime \prime \prime }\left( u_0\right) ,
\label{Ad3}
\end{equation}
\begin{equation}
A_{4}=u_{4}f^{\prime }\left( u_0\right) +\left[ \frac{1}{2!}%
u_{2}^{2}+u_{1}u_{3}\right] f^{\prime \prime }\left( u_0\right) +\frac{1}{2!}%
u_{1}^{2}u_{2}f^{\prime \prime \prime }\left( u_0\right) +\frac{1}{4!}%
u_{1}^{4}f^{(\mathrm{iv})}\left( u_0\right) .  \label{Ad4}
\end{equation}

Substituting Eqs. (\ref{7a}) and (\ref{8a}) into Eq. (\ref{6a}) we obtain
\begin{equation}
\mathcal{L}_t\left[ \sum_{n=0}^{\infty }u_{n}(t)\right] (s)=\frac{\beta
\left(N_1+N_2\right)/N-(\gamma +\mu)}{s(s+\mu)}-\frac{\mu \left[(\gamma
+\mu)-\beta \right] }{s^2(s+\mu)}-\frac{\left(\beta /N\right)\left[s+(\gamma
+\mu)\right]N_2}{s(s+\mu)}\mathcal{L}_t\left[\sum_{n=0}^{\infty }A_{n}\right]%
(s).  \label{11a}
\end{equation}

The matching of both sides of Eq. (\ref{11a}) gives the following iterative
algorithm for obtaining the power series solution of the basic evolution
equation of the SIR model with vital dynamics,
\begin{equation}  \label{12a}
\mathcal{L}_t\left[ u_{0}(t)\right](s) =\frac{\beta
\left(N_1+N_2\right)/N-(\gamma +\mu)}{s(s+\mu)}-\frac{\mu \left[(\gamma
+\mu)-\beta \right] }{s^2(s+\mu)},
\end{equation}%
\begin{equation}  \label{12b}
\mathcal{L}_t\left[ u_{1}(t)\right](s) =-\frac{\left(\beta/N\right)\left[%
s+(\gamma +\mu)\right]N_2}{s(s+\mu)}\mathcal{L}_t\left[A_{0}(t)\right](s) ,
\end{equation}%
\begin{equation}
\mathcal{L}_t\left[ u_{2}(t)\right](s) =-\frac{\left(\beta /N\right)\left[%
s+(\gamma +\mu)\right]N_2}{s(s+\mu)}\mathcal{L}_t\left[ A_{1}(t)\right](s) ,
\label{12c}
\end{equation}%
\begin{equation*}
...
\end{equation*}%
\begin{equation}
\mathcal{L}_t\left[ u_{k+1}(t)\right](s) =-\frac{\left(\beta /N\right)\left[%
s+(\gamma +\mu)\right]N_2}{s(s+\mu)}\mathcal{L}_t\left[ A_{k}(t)\right](s), k=0.1,2,... .
\label{12n}
\end{equation}

By applying the inverse Laplace transformation to Eq. (\ref{12a}), we obtain
the expression of $u_{0}(t)$ as
\begin{equation}
u_0(t)=\mathcal{L}_t^{-1}\left\{\frac{\beta \left(N_1+N_2\right)/N-(\gamma
+\mu)}{s(s+\mu)}-\frac{\mu \left[(\gamma +\mu)-\beta \right] }{s^2(s+\mu)}%
\right\}(t)=-\frac{\beta N_3}{\mu N }+\left( \beta -\gamma -\mu \right)t+\frac{%
\beta N_3 }{\mu N}e^{-\mu t}.
\end{equation}

We expand the term $\left(\beta N_3 /\mu N\right)e^{-\mu t}$ in power
series, and hence in the first order approximation we obtain
\begin{equation}
u_0(t)=\left[\frac{\beta \left(N_1+N_2\right)}{N}-\gamma -\mu\right]t.
\end{equation}

The first Adomian polynomial is given by
\begin{equation}\label{Ad0}
A_0(t)=e^{u_0(t)}=e^{\left[\frac{\beta \left(N_1+N_2\right)}{N}-\gamma -\mu%
\right]t}.
\end{equation}

Therefore, within the adopted approximation, we immediately obtain
\begin{eqnarray}
\hspace{-0.5cm} &&u_{1}(t)=-\mathcal{L}_{t}^{-1}\left\{ \frac{\left( \beta
/N\right) \left[ s+(\gamma +\mu )\right] N_{2}}{s(s+\mu )}\mathcal{L}_{t}%
\left[ e^{\left[ \frac{\beta \left( N_{1}+N_{2}\right) }{N}-\gamma -\mu %
\right] t}\right] (s)\right\} (t)=  \notag \\
\hspace{-0.5cm} &&\frac{\beta N_{2}}{\left( N_{1}+N_{2}\right) (\gamma
-\beta )+\gamma N_{3}}\left\{ \frac{\gamma }{\mu }-\frac{\beta
(N_{1}+N_{2})e^{\left[ \frac{\beta (N_{1}+N_{2})}{N}-\gamma \right] t}}{%
\left( N_{1}+N_{2}\right) (\gamma +\mu -\beta )+N_{3}(\gamma +\mu )}\right\}
e^{-\mu t}-\frac{\beta N_{2}(\gamma +\mu )}{\mu \left[ \left(
N_{1}+N_{2}\right) (\gamma +\mu -\beta )+N_{3}(\gamma +\mu )\right] }.
\end{eqnarray}%
The second Adomian polynomial is obtained as
\begin{equation}
A_{1}(t)=\frac{1}{1!}\frac{d}{d\lambda }\left. e^{\sum_{i=0}^{n}{%
u_{i}\lambda ^{i}}}\right\vert _{\lambda =0}=u_{1}(t)e^{u_{0}(t)},
\end{equation}%
and thus we find
\begin{eqnarray}\label{u2}
u_{2}(t) &=&-\mathcal{L}_{t}^{-1}\left\{ \frac{\left( \beta /N\right) \left[
s+(\gamma +\mu )\right] N_{2}}{s(s+\mu )}\mathcal{L}_{t}\left[ A_{1}(t)%
\right] (s)\right\} (t)=\frac{\beta ^{2}N_{2}^{2}}{2\left[ \left(
N_{1}+N_{2}\right) (\gamma +\mu -\beta )+N_{3}(\gamma +\mu )\right] ^{2}}%
\times  \notag \\
&&\Bigg\{ \frac{(\gamma +\mu )\left[ 2N(\gamma +\mu )-\beta \left(
N_{1}+N_{2}\right) \right] }{\mu \left[ (\gamma +2\mu )N-\beta \left(
N_{1}+N_{2}\right) \right] }-\frac{2\beta (\gamma +\mu )(N_{1}+N_{2})e^{%
\left[ \frac{\beta (N_{1}+N_{2})}{N}-\gamma -\mu \right] t}}{\mu \left[
\beta (N_{1}+N_{2})-\gamma N\right] }-  \notag \\
&&\frac{2\gamma e^{-\mu t}\left[ (2\gamma +\mu )N-\beta \left(
N_{1}+N_{2}\right) (\left( N_{1}+N_{2}\right) (\gamma +\mu -\beta
)+N_{3}(\gamma +\mu )\right] }{\mu \left[ \gamma N-\beta (N_{1}+N_{2})\right]
\left[ (2\gamma +\mu )N-2\beta (N_{1}+N_{2})\right] }-  \notag \\
&&\frac{2\gamma \left[ \beta (N_{1}+N_{2})-\mu N)(N_{1}(-\beta +\gamma +\mu
)+N_{2}(-\beta +\gamma +\mu )+N_{3}(\gamma +\mu )\right] e^{\left[ \frac{%
\beta (N_{1}+N_{2})}{N}-\gamma -2\mu \right] t}}{\mu \left[ \gamma N-\beta
(N_{1}+N_{2})\right] \left[ (\gamma +2\mu )N-\beta \left( N_{1}+N_{2}\right) %
\right] }+  \notag \\
&&\frac{\beta (N_{1}+N_{2})\left[ 2\beta \left( N_{1}+N_{2}\right) -(\gamma
+\mu )N\right] e^{2\left[ \frac{\beta (N_{1}+N_{2})}{N}-\gamma -\mu \right]
t}}{\left[ \beta (N_{1}+N_{2})-\gamma N)\right] \left[ 2\beta
(N_{1}+N_{2})-(2\gamma +\mu )N\right] }\Bigg\}.
\end{eqnarray}

Similarly, after computing the third Adomian polynomial, given by
\begin{equation}
A_2(t)=\frac{1}{2!}\frac{d^2}{d\lambda ^2}\left.e^{\sum _{i=0}^n{u_i\lambda
^i}}\right|_{\lambda =0},
\end{equation}
we obtain $u_3(t)$. The higher  order  terms in the Laplace-Adomian power
series representation of $u(t)$ can be obtained by following the same
approach, but due to their length we will not present them here. Hence, once
the expressions of the terms $u_n(t)$, $n=0,1,2,..$ in the Adomian
Decomposition are known, the approximate semianalytical solution of the SIR
model with vital dynamics can be obtained as
\begin{equation}
x(t)=\frac{1}{\left(\beta /N\right)}\sum _{n=0}^{\infty}{\frac{du_n(t)}{dt}}+%
\frac{\gamma +\mu}{\left(\beta /N\right)},
\end{equation}
\begin{equation}
y(t)=N_2e^{\sum _{n=0}^{\infty}{u_n(t)}},
\end{equation}
\begin{equation}
z(t)= e^{-\mu t}\left[N_3+\gamma N_2
{\int_0^t{e^{\mu t+\sum_{n=0}^{\infty}u_n(t)}dt}}\right]=e^{-\mu t}\left\{N_3+\gamma N_2\prod _{n=0}^{\infty}\int_0^t{e^{\mu t+u_n}dt}\right\}.
\end{equation}

The present Laplace-Adomian series solution is valid only for some specific ranges of the model parameters $(\beta, \gamma, \mu)$ and of initial conditions $\left(N_1,N_2,N_3\right)$. If $\left(N_1+N_2\right)N>\left(\gamma +\mu\right)/\beta$, or $\left(N_1+N_2\right)N>\left(\gamma +2\mu\right)/\beta$, the exponential functions in Eq.~(\ref{u2}) do diverge in the long time limit. Similarly, for $\beta /\gamma \rightarrow N/\left(N_1+N_2\right)$, $\beta /\left(\gamma +2\mu\right) \rightarrow N/\left(N_1+N_2\right)$, $2\beta \left(2\gamma +\mu\right)\rightarrow N/\left(N_1+N_2\right)$, singularities develop in the Adomian series.

\section{Comparison with the exact numerical solutions}\label{IV}

In the present Section we compare the semianalytical predictions of the Adomian and Laplace Adomian Decomposition Methods as applied to the SIR model with vital dynamics with the exact numerical solutions, obtained by numerically integrating the system of equations (\ref{K1})-(\ref{K3}).

\subsection{Comparison with the Adomian Decomposition Method}

In the case of the Adomian Decomposition Method we will approximate the solution by using only two first two terms in the Adomian iterative scheme.  However, we will also add to the series solution a truncation of the term $u_2$, approximated by $u_2(t)\approx -\mu \int_0^t{u_1(t)dt}$. Hence we approximate the Adomian solution of the basic evolution equation of the SIR model with vital dynamics as
\begin{eqnarray}
\hspace{-0.5cm}&&u(t)\approx   \nonumber \\
\hspace{-0.5cm}&&\frac{3\sqrt{\pi }\beta N_{2}e^{\frac{a^{2}}{4b}}\left[ \text{erf}\left(
\frac{a}{2\sqrt{b}}\right) -\text{erf}\left( \frac{a+2bt}{2\sqrt{b}}\right) %
\right] \left[ a(\gamma +\mu )+2b(t(\gamma +\mu )+1)\right] +2\sqrt{b}\left[
3\beta N_{2}(\gamma +\mu )\left( 1-e^{-t(a+bt)}\right) +b\mu Nt^{2}(3a+2bt)%
\right] }{12b^{3/2}N}  \nonumber \\
\hspace{-0.5cm}&&-\frac{\mu}{48b^{5/2}N} \Bigg\{3\sqrt{\pi }\beta N_{2}e^{\frac{a^{2}}{4b}}%
\left[ \text{erf}\left( \frac{a}{2\sqrt{b}}\right) -\text{erf}\left( \frac{%
a+2bt}{2\sqrt{b}}\right) \right] \left[ a^{2}(\gamma +\mu )+4a(bt(\gamma
+\mu )+b)+2b(2bt(t(\gamma +\mu )+2)+\gamma +\mu )\right] \nonumber\\
\hspace{-0.5cm}&&+2\sqrt{b}\left[
4ab^{2}\mu Nt^{3}-3\beta N_{2}e^{-t(a+bt)}(a(\gamma +\mu )+2b(t(\gamma +\mu
)+2))+3a\beta N_{2}(\gamma +\mu )+2b^{3}\mu Nt^{4}+12b\beta N_{2}(t(\gamma
+\mu )+1)\right] \Bigg\}\nonumber\\
\hspace{-0.5cm}&&-at-bt^{2}.
\end{eqnarray}

\begin{figure*}[h]
\centering
\includegraphics[scale=0.7]{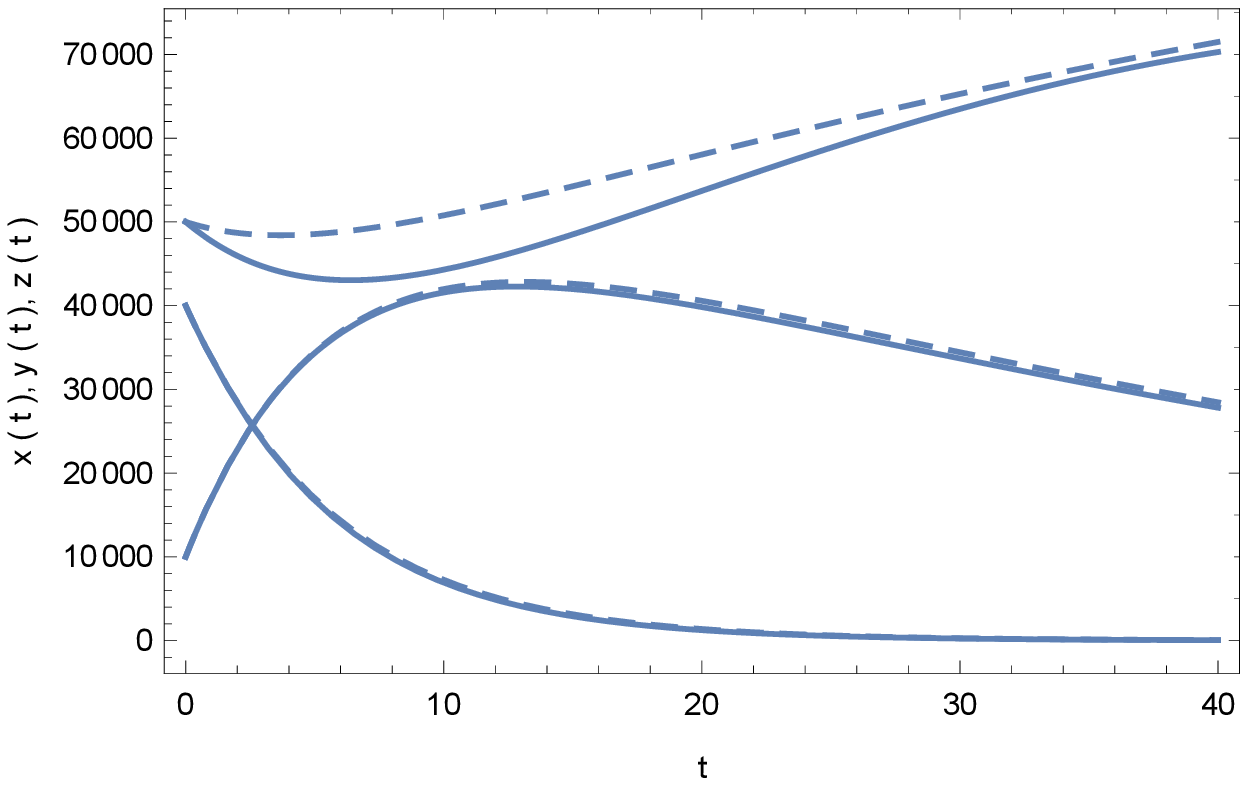}
\includegraphics[scale=0.7]{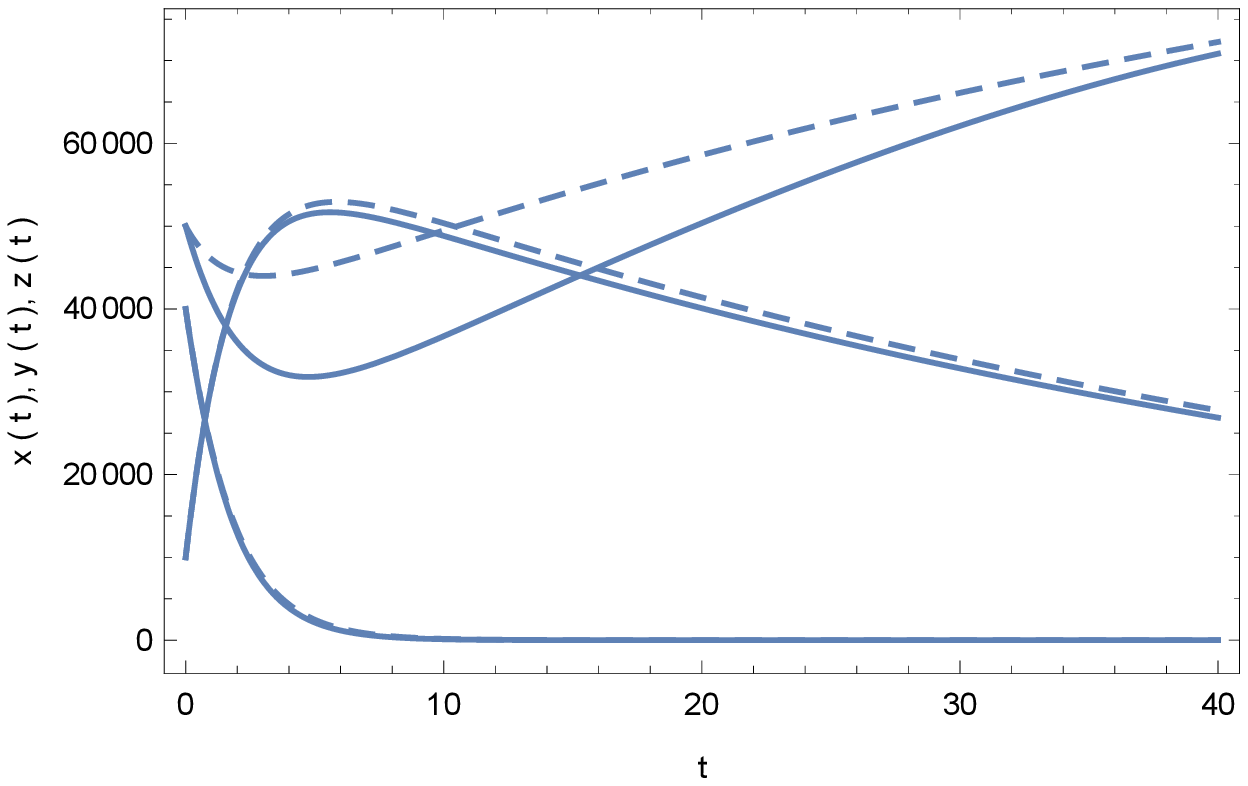}
\caption{Comparison of the exact numerical solution of the SIR model with vital dynamics and of the semianalytical solution obtained with the use of the Adomian Decomposition Method, for $\beta =0.10$, $\gamma =0.20$ and $\mu =0.02$ (left panel), and $\beta =0.34$, $\gamma =0.69$ and $\mu =0.02$ (right panel), respectively. The results of the numerical integration are represented by the dashed curves, while the semianalytical approximations are represented by the solid curves. The initial conditions used to  numerically integrate the system of Eqs.~(\ref{K1})-(\ref{K3}) are $x(0)=N_1=50000$, $y(0)=N_2=40000$ and $z(0)=N_3=10000$.}
\label{fig1}
\end{figure*}

The comparison of the numerical and of the semianalytical Adomian approximate solution is presented, for different values of the model parameters $\beta$, $\gamma $ and $\mu$ in Fig.~\ref{fig1}.

\subsection{Comparison with the Laplace-Adomian Decomposition Method}

\begin{figure*}[h]
\centering
\includegraphics[scale=0.7]{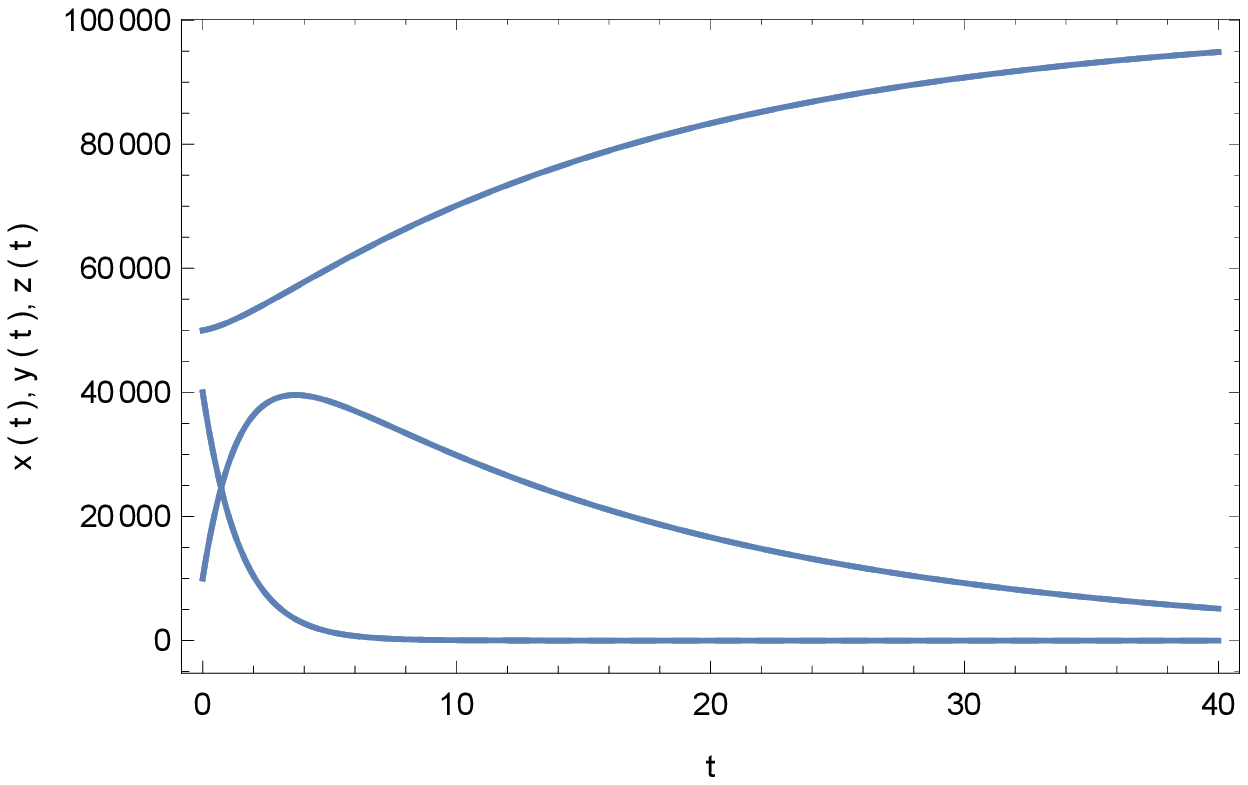}
\includegraphics[scale=0.7]{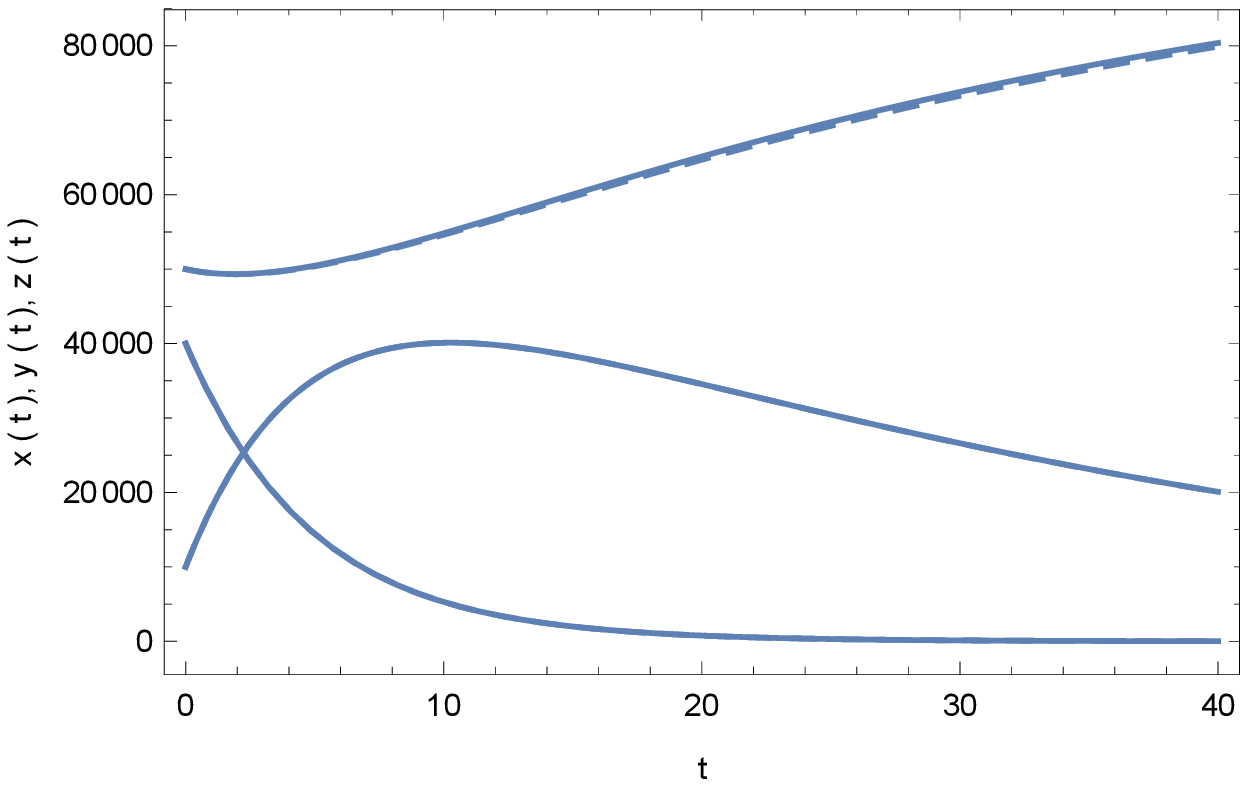}
\caption{Comparison of the exact numerical solution of the SIR model with vital dynamics and of the semianalytical solution obtained with the use of the Laplace-Adomian Decomposition Method, for $\beta =0.11$, $\gamma =0.67$ and $\mu =0.05868$ (left panel), and $\beta =0.11$, $\gamma =0.23$ and $\mu =0.02868$ (right panel), respectively. The results of the numerical integration are represented by the dashed curves, while the semianalytical approximations are represented by the solid curves. The initial conditions used to  numerically integrate the system of Eqs.~(\ref{K1})-(\ref{K3}) are $x(0)=N_1=50000$, $y(0)=N_2=40000$ and $z(0)=N_3=10000$.}
\label{fig2}
\end{figure*}

\begin{figure*}[h]
\centering
\includegraphics[scale=0.7]{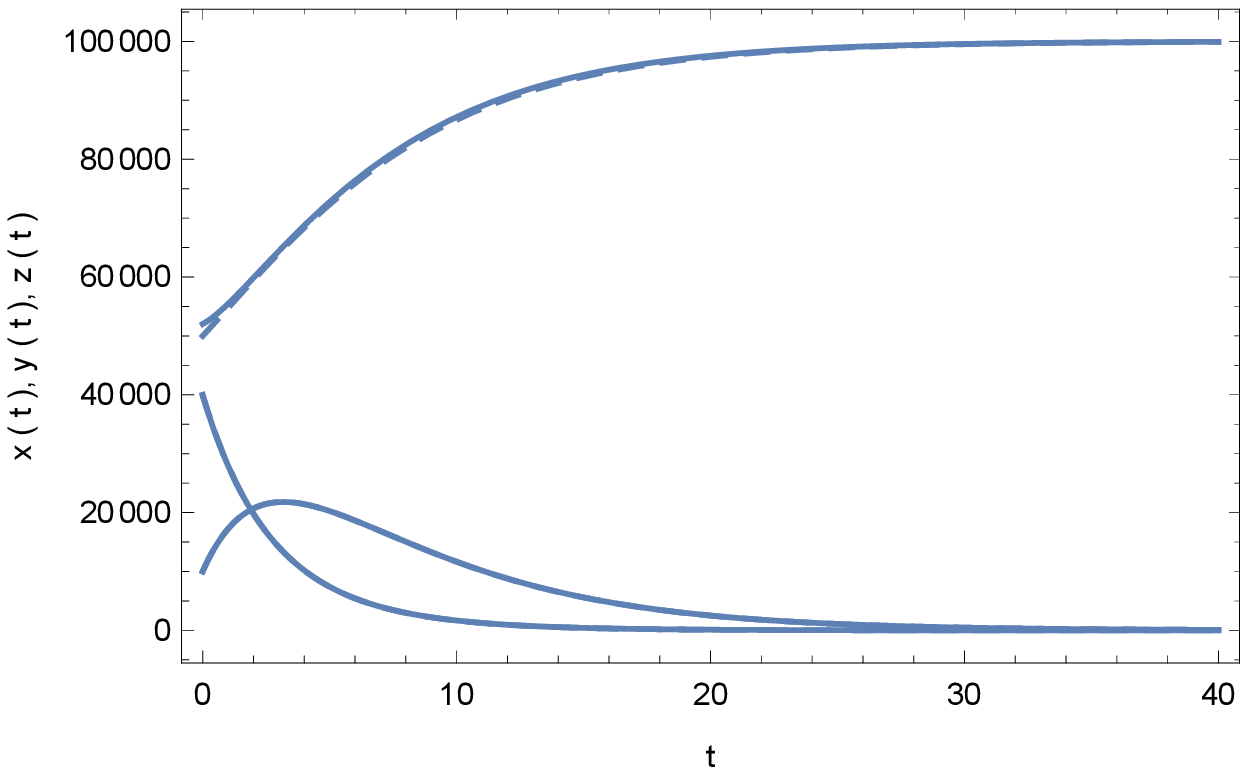}
\includegraphics[scale=0.7]{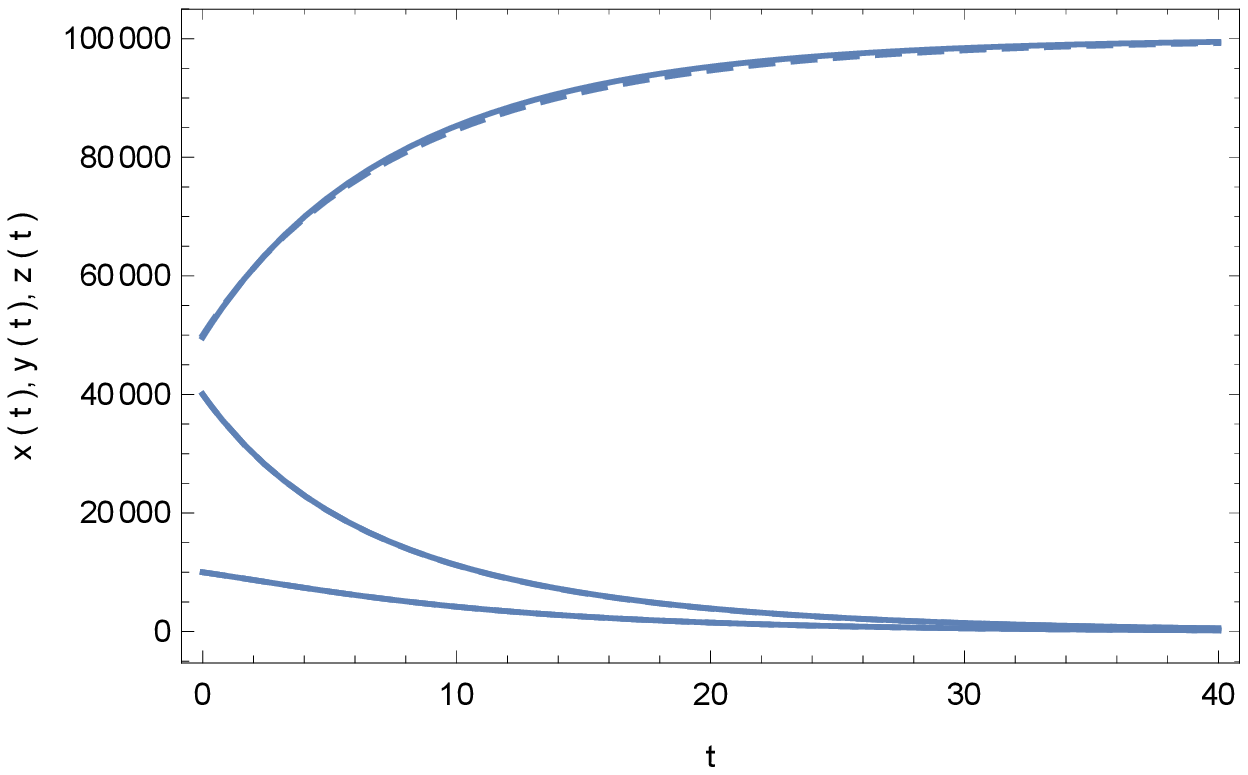}
\caption{Comparison of the exact numerical solution of the SIR model with vital dynamics and of the semianalytical solution obtained with the use of the Laplace-Adomian Decomposition Method, for $\beta =0.11$, $\gamma =0.67$ and $\mu =0.05868$ (left panel), and $\beta =0.11$, $\gamma =0.029$ and $\mu =0.177$ (right panel), respectively. The results of the numerical integration are represented by the dashed curves, while the semianalytical approximations are represented by the solid curves. The initial conditions used to  numerically integrate the system of Eqs.~(\ref{K1})-(\ref{K3}) are $x(0)=N_1=50000$, $y(0)=N_2=40000$ and $z(0)=N_3=10000$.}
\label{fig3}
\end{figure*}

The Laplace-Adomian Decomposition Method, as applied in the present version to the analysis of the SIR model with vital dynamics allows the computation of an arbitrary number of terms in the Adomian series expansion. The comparison of the exact numerical results with the Laplace-Adomian series containing six terms, is presented, for a selected number of values of the model parameters $(\beta, \gamma, \mu)$, in Figs.~\ref{fig2} and \ref{fig3}, respectively.

As one can see from the Figures, for the adopted values of the model parameters there is a good concordance between the numerical and the semianalytical results. Of course increasing the number of terms of the Laplace-Adomian series would increase the precision of the results. On the other hand, depending the necessary accuracy in the investigated problem, a smaller number of terms may offer enough precision in the description, analysis and interpretation of the epidemiological data. For a small number of terms explicit analytical representations of the solution can be obtained in a simple form.

\section{Discussions and final remarks}\label{V}

As of the moment of writing of the present paper the number of individuals infected with the Covid-19 virus has exceeded 24 millions, with the number of fatalities higher than 800,000. There is no sign yet for a slowing down of the epidemics, and second or third waves ar still hitting different regions of the planet. Despite that the fight against the virus, and finding a cure (or a vaccine) for it is mostly a medical/biological/virusological problem, the understanding of the spread of the epidemics may lead to the adoption/imposition of quarantine or safety measures that could drastically reduce its intensity. In this context the mathematical epidemiological models could play an important role, since a successful modeling of the spread of the disease could essentially contribute, on the level of the society, to the implementation of the best policies that could guarantee the maximal safety of the citizens.

Despite their (apparent) mathematical simplicity, the compartmental epidemiological models did play an important role in the analysis of the Covid pandemic. These models contain the basic features of the evolution of an epidemics, and once the parameters of the model are fixed from the epidemiological data they can provide some accurate predictions for the evolution of the infectious diseases. From a mathematical point of view the SIR model is exactly integrable, which simplifies its analysis, since its exact solution is known \cite{Harko1}. On the other hand the SIR model with vital dynamics is non-integrable, and therefore it can be investigated only by using numerical or semianalytical methods. In the present paper we have introduced the powerful Adomian and Laplace-Adomian Decomposition Methods for the study of this model. However, in obtaining the explicit form of the Adomian polynomials one must use some approximations for their estimation. In the present approach we have approximated the $u_0(t)$ term by its first order series expansion, leading to the first Adomian polynomial as given by Eq.~(\ref{Ad0}). In the large time limit if $\beta \left(N_1+N_2\right)/N>\gamma +\mu $, $A_0$ diverges. Hence the Laplace-Adomian Decomposition Method works in the present case optimally if the condition $\left(N_1+N_2\right)/N<\left(\gamma +\mu\right)/\beta$.

In the present paper we have presented a series solution of the  non-integrable SIR epidemiological model with vital dynamics, by using both the Adomian and the Laplace-Adomian Decomposition Methods. The application of these methods allows to obtain
the explicit time dependencies of $x(t)$, $y(t)$ and $z(t)$. The semianalytical solutions have a simple mathematical form, and they
describes very precisely the numerical behavior of the model for a large range of the model parameters $(\beta, \gamma, \mu)$. The time dependence of
the three compartments in the semianalytical solution obtained by using the Laplace-Adomian Decomposition Method is given by a sum containing exponential terms. Such a representation may simplify the fitting with the epidemiological results. The series, truncated to a small number of terms, give a very
good description of the numerical results for a large number of values of the model parameters, and of the initial conditions. In fact, the two
terms approximation obtained with the use of the Adomian Decomposition Method also gives a good approximation of the results of the numerical integration of the SIR model with vital dynamics for a large range of parameter values.

Exact solutions of the epidemiological models are important for epidemiologists because they allow the study of
the spreading of infectious diseases in different situations. They are also helpful in the design of the best social strategies for their control.
Hopefully the results obtained in the present paper may also contribute to the investigations of the dynamics, evolution and long term impact of the
present and of the future epidemics.

%\section*{Acknowledgments}

\end{document}